# Preparation of Carbon Nanotube Bioconjugates for Biomedical Applications

Zhuang Liu,[1,2] Scott Tabakman,[1] Zhuo Chen,[1] Hongjie Dai[1]

[1]Department of Chemistry, Stanford University, Stanford, CA, USA, [2]Functional Nano & Soft Materials Laboratory (FUNSOM), Soochow University, Suzhou, Jiangsu, China

Correspondence should be addressed to Z.L (zliu@suda.edu.cn) or H.D (hdai@stanford.edu)

**Abstract**

Biomedical applications of carbon nanotubes have attracted much attention in recent years. Here, we summarize our previously developed protocols for functionalization and bioconjugation of single wall carbon nanotubes (SWNTs) for various biomedical applications including biological imaging, sensing and drug delivery. Sonication of SWNTs in solutions of phospholipid-polyethylene glycol (PL-PEG) is our most commonly used protocol of SWNT functionalization. Compared to other frequently used covalent strategies, our non-covalent functionalization protocol largely retains the intrinsic optical properties of SWNTs, which are useful in various biological imaging and sensing applications. Functionalized SWNTs are conjugated with targeting ligands for specific cell labeling *in vitro* or tumor targeting *in vivo*. Radio labels are introduced for tracking and imaging of SWNTs in real time *in vivo*. Moreover, SWNTs can be conjugated with small interfering RNA (siRNA) or loaded with chemotherapy drugs for drug delivery. These procedures take various times ranging from one to five days.

# INTRODUCTION

Carbon nanotubes with various unique physical and chemical properties have shown interesting applications in many fields including biomedicine.[1, 2] Functionalized carbon nanotubes with water solubility and biocompatibility are able to cross cell membranes, shuttling a wide range of biologically active molecules including drugs, proteins, DNA and RNA, into cells.[3-7] The cytotoxicity of carbon nanotubes is largely dependent on their surface functionalization, with minimal toxic effects for well functionalized, serum-stable nanotubes.[1, 8] We have shown that after intravenous injection into mice, well functionalized single-walled carbon nanotubes (SWNTs) are accumulated in reticuloendothelial systems (RES) are slowly excreted, mainly through the biliary pathway, without exhibiting obvious side effects to the treated mice[9, 10]. Recently, *in vivo* cancer treatment in an animal model has been realized by carbon nanotube-based drug delivery.[11]

Carbon nanotubes are classified as single-walled carbon nanotubes (SWNTs) and multi-walled carbon nanotubes (MWNTs), depending on the number of graphene layers from which a nanotube is composed. SWNTs are quasi one dimensional quantum wires with sharp densities of electronic states (electronic DOS) at the van Hove singularities, and generally have more attractive unique intrinsic optical properties than MWNTS. SWNTs can be used as optical tags for biomedical detection and imaging.[12-18] Ultra-sensitive *ex vivo* protein sensing with a detection limit as low as one femtomolar using SWNT Raman tags has been achieved utilizing the resonance Raman scattering property of SWNTs and surface enhanced Raman scattering (SERS).[14] The Raman scattering, near-infrared (NIR) photoluminescence, and high optical absorbance of SWNTs have all been utilized for biomedical molecular imaging *in vitro* and *in vivo*.[15-18] Thus, carbon nanotube bioconjugates are promising nanomaterials for biomedical applications.

SWNTs may have various potential advantages over other nanomaterials in different areas of nanobiotechnology. As an optical tag in biological imaging, SWNTs can be used in Raman, NIR fluorescence and photoacoustic imaging of cells and animals.[12, 13, 15-18] Multimodality optical imaging could thus be achieved using SWNTs as the contrast agent. Quantum dots or surface enhanced Raman scattering gold / sliver nanoparticles normally only have a single imaging functionality.[19, 20] In contrast to widely used fluorescent quantum dots, carbon nanotubes contain no heavy metals and thus have a safer chemical composition. In the area of drug delivery, SWNT-based siRNA delivery works for a large range of cells, including notoriously 'hard-to-transfect' human T cells,[4] which are inert to conventional liposomal transfection agents. SWNTs can be efficiently loaded with aromatic chemotherapy drugs such as doxorubicin via supramolecular π-π stacking,[21] obtaining an ultra-high loading capacity superior to other drug carriers including liposomes and micelles. Moreover, the high optical absorbance of SWNTs can also be used in photothermal therapy,[22, 23] which may potentially be combined with chemotherapy[11, 21] and gene therapy,[4, 24] both delivered by SWNTs to treat cancer in a more efficient manner.

For biomedical applications, raw hydrophobic carbon nanotubes must be functionalized to afford water solubility and biocompatibility. Our previous studies have uncovered that the behaviors of carbon nanotubes in biological systems *in vitro* (such as cellular uptake) and *in vivo* (such as blood circulation time and biodistribution) are highly dependent on their surface chemistry.[4, 9, 25] Developing proper surface functionalization on carbon nanotubes is thus the most critical step to produce nanotube bioconjugates for a desired application. There are two major types of functionalization protocols for carbon nanotubes, covalent reactions or non-covalent coating by amphiphilic molecules on nanotubes. Various covalent functionalization reactions, such as oxidation[26, 27] of nanotubes and 1, 3-dipolar cylcoaddtion[28] on the nanotube sidewalls, have been developed to produce water soluble nanotubes useful in certain biomedical applications such as drug delivery.[2] Although covalent chemical reactions often allow stable functionalization on carbon nanotubes, the properties of SWNTs are degraded when the nanotube side wall is damaged, dramatically decreasing the Raman scattering and NIR fluorescence signals of SWNTs.[1] Therefore, covalently functionalized carbon nanotubes, have been widely used in drug and gene delivery,[2, 29] but are usually not ideal for sensing and imaging applications.[1] In contrast, the structure and optical properties of SWNTs are largely maintained when non-covalent functionalization is used. However, the stability and biocompatibility of many non-covalently functionalized SWNTs are not satisfactory. For example, SWNTs solubilized in small molecule surfactants (e.g., sodium dodecyl sulfate, SDS) will aggregate and precipitate if excess coating molecules are removed. An ideal functionalization should impart SWNTs with high water solubility, biocompatibility, minimal damage of nanotube structure, and functional groups available for further bioconjugation.

Our group has developed systematic protocols for SWNT functionalization and bioconjugation in the past few years. Raw SWNTs are non-covalently functionalized by amphiphilic polymers, such as phospholipid-poly(ethylene glycol) (PL-PEG).[6, 22] Functionalized SWNTs have excellent stability in the aqueous phase and are highly biocompatible. Targeting ligands including antibodies and peptides can be conjugated to SWNTs to recognize specific cell receptors, yielding targeted SWNT bioconjugates useful for biological sensing[14] and imaging.[15-18] We have also developed a protocol to label SWNTs with radioactive isotopes to track and image nanotubes *in vivo* by positron emission tomography (PET). In addition, SWNT-based siRNA transfection can be achieved by conjugating siRNA to SWNTs via a cleavable disulfide bond.[4, 6] Furthermore, aromatic drug molecules can be non-covalently loaded on SWNTs by simple mixing for drug delivery.[21]

Here, we systematically summarize the nanotube functionalization and bioconjugation protocols developed and used in our previous studies. Although our bioconjugation strategies apply for a wide range of biomolecules, only a few model systems are chosen to illustrate those protocols. Arg–Gly–Asp (RGD) peptide and Herceptin anti-Her2 antibody are used as targeting ligands. $^{64}$Cu is reported as an example of radiolabeling SWNTs. Anti-CXCR4 siRNA is chosen for siRNA conjugation and delivery. Finally, doxorubicin is demonstrated as an aromatic drug, loaded onto SWNTs for drug delivery. These detailed protocols should be beneficial to scientists interested in further developing biological applications of novel nanomaterials.

**Experimental design**

**Non-covalent Functionalization of SWNTs by PL-PEG**

SWNTs are non-covalently functionalized by sonication of raw, hydrophobic nanotubes in water solutions of amphiphilic polymers (e.g., PL-PEG).[6, 22] The hydrophobic lipid chains of PL-PEG are strongly anchored onto the nanotube surface while the hydrophilic PEG chain affords SWNT water solubility and biocompatibility. After removal of excess PL-PEG molecules, functionalized SWNTs exhibit excellent stability in various aqueous phases including water, physiological buffers (e.g., phosphate buffered saline, PBS), cell medium and whole serum. The concentration of a SWNT solution can be determined by its optical density at 808 nm measured by a UV-VIS-NIR spectrometer with a weight extinction coefficient of 0.0465 mg L$^{-1}$ cm$^{-1}$ (dividing the optical density at 808 nm by the extinction coefficient gives the concentration) [22] The length distribution of functionalized SWNTs can be determined by an atomic force microscope (AFM). Those non-covalently functionalized SWNTs retain their Raman and NIR fluorescence properties and are useful in biological detection and imaging applications. The functional group (e.g., amine) on the PEG terminal is available for further bioconjugation (Fig. 1).

**Conjugation with Targeting Ligands**

Targeting ligands including antibodies and peptides can be conjugated to SWNTs to recognize specific cell receptors (Fig., 2). Herceptin is a monoclonal antibody that binds specifically to the Her2/neu receptor over-expressed on a wide range of human breast cancer cells.[30] The RGD peptide targets integrin $\alpha_v\beta_3$ receptors that are up-regulated on fast-growing tumor vasculature cells and many types of cancer cells.[31] The antibody Herceptin is first thiolated by Traut's reagent following standard protocols[15, 16] and used immediately after purification. The Traut's reagent reacts with amino groups on the antibody and produces active thiol groups useful for bioconjugation. Thiolated RGD peptide synthesized following a previously published protocol [32] is used directly. The thiolated antibody or peptide should be protected from oxidation by adding Ethylenediaminetetraacetic acid (EDTA) to prevent heavy metal catalyzed oxidization, or Tris- (2-carboxyethyl)phosphine hydrochloride (TCEP) as a reducing agent, during the conjugation with nanotubes. Maleimide groups are introduced onto SWNTs by reacting PL-PEG-amine functionalized SWNTs with a sulfosuccinimidyl 4-N-maleimidomethyl cyclohexane-1-carboxylate (Sulfo-SMCC) bifunctional linker. The activated SWNTs are then reacted with thiolated antibodies or peptides, obtaining targeted SWNT bioconjugates, which can be used in multiplexed Raman spectroscopic imaging,[15, 18] ultra-sensitive Raman detection of proteins,[14] NIR fluorescence imaging,[16] photoacoustic imaging,[17] and targeted photothermal therapy.[22, 23] The targeting ability of SWNT bioconjugates (e.g., SWNT-RGD) can be characterized by *in vitro* Raman spectroscopic imaging experiments, to examine the staining of integrin $\alpha_v\beta_3$ positive U87MG cells and $\alpha_v\beta_3$ negative MCF-7 cells incubated with the SWNT-RGD conjugate .

**Radiolabeling of SWNTs**

To image and track SWNTs *in vivo* by positron emission tomography (PET), SWNTs are labeled with a radioactive isotope (Fig. 3).[25] PET imaging provides three dimensional distribution information of radio labeled nanotubes in live mice at the real time. To obtain RGD conjugated radiolabeled SWNTs, SWNTs are first reacted with a mixture of sulfo-SMCC and N-Hydroxysuccinimide (NHS) activated 1,4,7,10-Tetraazacyclododecane-1,4,7,10-tetraacetic acid (DOTA), and then conjugated to RGD-SH. After removal of excess reagents, $^{64}$Cu radioactive isotope can then be complexed to the DOTA rings on the SWNTs to achieve radiolabeling. The radiolabeled, targeted SWNT bioconjugate can be used for *in vivo* PET imaging of mice bearing integrin $\alpha_v\beta_3$ positive e.g., U87MG human glioblastomas tumors.[25] 5-10 million of U87MG cells should be injected subcutaneously on the shoulder of a nude mouse. The mice can be used 2-3 weeks after tumor inoculation. PET imaging should be carried out at 0.5 h, 2 h, 4 h, 6 h and 24 h post injection (p.i.). Mice may be sacrificed at 24 h p.i. when the blood circulation of nanotubes is ended.

**SiRNA Conjugation to SWNTs via Cleavable Disulfide Bond**

The intracellular molecular delivery ability of SWNTs can be used for siRNA transfection.[4, 6] In this example we chose to use CXCR4, a chemokine receptor that plays an important role in the entry of HIV virus into human T cells.[33] SWNTs are first reacted with a bifunctional linker, *N*-Succinimidyl 3-(2-pyridyldithio)-propionate (Sulfo-LC-SPDP), and then conjugated with thiolated siRNA via a cleavable disulfide bond (Fig. 4). Two CXCR4 siRNA with different sequences and a control luciferase siRNA are used. Dithiothreitol (DTT) is used to cleave any disulfide bond formed during storage of thiolated siRNA and removed by a NAP-5 column prior to the conjugation of siRNA with SWNTs. The SWNT-siRNA conjugates should be sterilized by centrifugation before cell incubation. Once transported into cells via endocytosis, siRNA is released from SWNTs by sulfide cleavage and then binds to CXCR4 mRNA to induce gene silencing. CXCR4 receptor expression on CEM.NK$^R$ cells, a human T cell line inert to commercial cationic liposome transfection agents,[34] is knocked-down after cells are incubated with the SWNT-siRNA$_{anti-CXCR4}$ conjugate for 3 days. The CXCR4 expression levels of cells can be determined by labeling cells with PE-anti-CXCR4 antibody and flow cytometry (FACS) measurement with dead cells excluded by propidium iodide (PI) staining. Control experiments using commercial cationic liposome based transfection agents to transfect CEM.NK$^R$ with CXCR4 siRNA showed no obvious gene silencing effect. Mis-matched siRNA sequence (luciferase siRNA) also did not affect CXCR4 expression.

**Doxorubicin Loading on Functionalized SWNTs**

SWNTs, with all atoms exposed on their surface, have ultra-high surface area available for binding of aromatic molecules via supramolecular π-π stacking.[21] Functionalized SWNTs with or without targeting ligands can be loaded with doxorubicin (DOX), an aromatic chemotherapy drug used for various types of cancers, by simply mixing of the two solutions at a slightly basic pH. Excess un-loaded DOX can be removed by filtration. The optical absorption of SWNT-DOX at 490 nm after subtraction of the SWNT absorption (at the same nanotube concentration) can be used to calculate the DOX concentration and loading in the SWNT-DOX complex.[21] Based on the UV-VIS-NIR absorption spectra, up to 4 grams of DOX can be loaded on 1 gram of SWNTs. The toxicity of SWNT-DOX is lower than that of free DOX, but can be enhanced when conjugated with a targeting ligand such as RGD peptide, for targeted drug delivery. Toxicity assays are based on triplicate sample and can be carried out by testing cell viabilities after incubating cells with free DOX, SWNT-DOX and SWNT-RGD-DOX at series of DOX concentrations using a CellTiter 96 kit (Promega). The cell viability of the untreated control is defined as 100%.

## MATERIALS

**Reagents**

- Hipco single walled carbon nanotubes (Unidym Inc)

**! CAUTION** Hipco SWNTs are very light and could become airborne. Inhalation of SWNTs should be avoided. Wear goggles, lab coat and face-mask during experiments.

- Isopotically Modified pure C12, pure C13 and mixed C12/C13 SWNTs, (Tsinghua-Foxconn Nanotechnology Research Center, China; see Reagent Setup)
- PL-PEG5000-Amine (NOF cooperation, cat. no. DSPE-050PA)
- PL-PEG2000-Amine (NOF cooperation, cat. no. DSPE-020PA)
- RGD peptide (c(RGDyK); Peptides International)
- Thiolated RGD peptide (see Reagent Setup)
- Herceptin (Genetech Inc)
- Sulfo-SMCC (Sulfosuccinimidyl 4-N-maleimidomethyl cyclohexane-1-carboxylate) (Pierce,

cat. no. 22322)

- Sulfo-NHS (*N*-hydroxysulfosuccinimide) (Pierce, cat. no. 24510)
- Sulfo-LC-SPDP (*N*-Succinimidyl 3-(2-pyridyldithio)-propionate) (Pierce, cat. no. 21650)
- Traut's Reagent (2-Iminothiolane•HCl) (Pierce, cat. no. 26101)
- EDC (1-Ethyl-3-[3-dimethylaminopropyl]carbodiimide hydrochloride) (Pierce, cat. no. 22981)
- DOTA (1,4,7,10-Tetraazacyclododecane-1,4,7,10-tetraacetic acid) (Sigma-Aldrich, cat. no. 86734)
- DTT (DL-Dithiothreitol) (Sigma-Aldrich, cat. no. 43815)
- Sodium Bicarbonate (Sigma-Aldrich, cat. no. S6297)
- Sodium Hydroxide (Sigma-Aldrich, cat. no. 221465)

**! CAUTION** Sodium Hydroxide is a corrosive strong base. Wear goggles, lab coat and face-mask during experiments.

- DMSO (Dimethyl sulfoxide), 99.9%, anhydrous (Sigma-Aldrich, cat. no. 276855)
- EDTA (Ethylenediaminetetraacetic acid) solution (Sigma-Aldrich, cat. no. 03690)
- TCEP (Tris(2-carboxyethyl)phosphine hydrochloride) (Sigma-Aldrich, cat. no. C4706)
- Sodium Acetate (Sigma-Aldrich, cat. no. S8750)
- Acetic Acid (Sigma-Aldrich, cat. no. 242853)

**! CAUTION** Acetic Acid is evaporative and corrosive. Wear goggles, lab coat and face-mask during experiments. Handle acetic acid inside a hood.

- Chelex 100 sodium form (Sigma-Alrich, cat. no. C7901)
- DOX (Doxorubicin HCl, Tecoland)
- U87MG human glioblastoma cells (ATCC, cat. no. HTB-14; see Reagent Setup)
- MCF-7 human breast cancer cells (ATCC, cat. no. HTB-22; see Reagent Setup)
- BT-474 human breast cancer cells (ATCC, cat. no. HTB-20; see Reagent Setup)
- CEM.NK$^R$ human T-lymphoblastoid cells (NIH Aids Reagents Program, cat. no. 795081; see Reagent Setup)

- Low-glucose DMEM Medium (Invitrogen, cat. no. 11885-084)
- High-glucose DMEM Medium (Invitrogen, cat. no. 21063-045)
- RMPI-1640 Medium (Invitrogen, cat. no. 11875-093)
- Fetal bovine serum (FBS) (Invitrogen, cat. no. 10437-028)
- Phosphate buffered saline (PBS), 10×, pH 7.4 (Invitrogen, cat. no. 70011-069)
- Phosphate buffered saline (PBS), 1×, pH 7.4 (Invitrogen, cat. no. 10010-049)
- Penicillin-Streptomycin, liquid (10,000 units penicillin;10,000 μg streptomycin) (Invitrogen cat. no. 15140-163)
- CXCR4 siRNA: sequence-*a*: 5'-Thiol-GCG GCA GCA GGU AGC AAA GdTdT-3'; sequence-*b*: 5'-Thiol-AUG GAG GGG AUC AGU AUA UdTdT. (Dharmacon RNAi Technology)
- Luciferase siRNA (control): 5'-Thiol-CUU ACG CUG AGU ACU UCG AdTdT-3'. (Dharmacon RNAi Technology)
- PE-anti-CXCR4 antibody, (Caltag Laboratories, cat. no. MHCXCR404)
- Propidium iodide (PI) solution, 1.0 mg/ml in water (Sigma-Aldrich, cat. no. P4864)

**! CAUTION** Propidium iodide causes eye, skin and respiratory irritation and is harmful if swallowed. Wear goggles, lab coat and face-mask during experiments.

- Lipofectamine2000 (Invitrogen, cat. no. 11668019)
- LipofectamineRNAiMAX (Invitrogen, cat. no. 13778075)
- siPORT (Ambion, cat. no. AM4510)
- HiPerFect (Qiagen, cat. no. 301704 )
- CellTiter 96 MTS assay kit (Promega, cat. no. G3580)
- Isoflurane (RxElite Inc., cat. no. NDC60307-120-25)

**! CAUTION** Isoflurane is a profound respiratory depressant. Wear goggles, lab coat and face-mask during experiments. Handle isoflurane inside a hood when appropriate. Closely seal the bottle after use.

- $^{64}CuCl_2$ radioactive isotope (University of Wisconsin-Madison)

**! CAUTION** Please obtain appropriate training for handling radioactive materials. Wear goggles, lab coat, mask, radiation dosimeter badge and rings during experiment. Handling of radioactive isotopes should only be performed in designated rooms for those experiments. Check any possible radioactive contamination after experiments.

- U87MG tumor-bearing athymic nude mice (Harlan; see Reagent Setup)

**! CAUTION** Please obtain appropriate training from the institution regarding animal handling. Animal protocols must be in place before performing animal studies.

**Equipment**

- Bath Sonicator (Cole-Parmer, cat. no. 08849-00)
- Accu Spin 400 centrifuge (Fisher Scientific, cat. no. 75005194)
- Sorvall Legend Mach 1.6R centrifuge (Kendro Laboratory Products, cat. no. 75004337)
- Amicon centrifugal filter device 4ML, 10,000 MWCO (Millipore, cat. no. UFC801024)
- Amicon centrifugal filter device 4ML, 100,000 MWCO (Millipore, cat. no. UFC810024)
- Microcon Ultracel YM-100 filter device 0.5ML, 100,000 MWCO (Millipore, cat. no. 42412)
- Illustra NAP-5 columns, Sephadex G-25 DNA Grade (GE Healthcare, cat. no. 17-0853-01)
- Flow cytometer (FACScan, Becton Dickinson)
- Cary-6000i UV-VIS-NIR Spectrometer (Varian Inc)
- Tecan Spectrafluor Plus microplate reader (Tecan Group Ltd)
- Confocal Raman microscope (Horiba-Jobin-Yvon Inc)
- Micro-PET R4 rodent model scanner (Concorde Microsystems)

**Reagent Setup**

**Isotopically Modified SWNTs**   $^{13}$C-doped SWNTs are produced by Profs. Kaili Jiang and Shoushan Fan at Tsinghua-Foxconn Nanotechnology Research Center, Department of Physics, Tsinghua University, China. The detailed method has been reported earlier.[15]

**Thiolated RGD (RGD-SH) Peptide**   Thiolated RGD (RGD-SH) peptide is prepared following

a previously reported protocol.[32]

**10 mM RGD-SH solution**   Dissolve 1.2 mg of RGD-SH in 200 μl water (~10 mM). Store RGD-SH solution at -20 °C in small aliquots to avoid too many freeze-thaw cycles. The solution can be stable for up to 6 months if used and stored properly.

**100 μM SiRNA solution**   Dissolve siRNA purchased from Dharmacon in the desired amount of RNAse free water to reach a siRNA concentration of 100 μM. Store siRNA solution at -20 °C in small aliquots to avoid too many freeze-thaw cycles. The solution can be stable for up to 6 months if used and stored properly.

**0.5 M Sodium Bicarbonate ($NaHCO_3$) Buffer**   Dissolve 8.4 g sodium bicarbonate in 200 ml water. The solution can be stored at room temperature in a plastic bottle and stable for 6 months.

**0.1 M Sodium Hydroxide (NaOH) Solution**   Dissolve 1.0 g sodium hydroxide in 250 ml water. The solution can be stored at room temperature in a plastic bottle and stable for 6 months.

**0.1 M Sodium Acetate Buffer (NaAcO, pH6.5)**   Dissolve 1.64 g sodium acetate in 200 ml water. Add 22 mg acetic acid liquid. Add 1 g of Chelex 100 beads into the buffer to avoid heavy metal ion contamination.

**Cell Culture**   Culture U87MG cells in DMEM (low glucose) supplemented with 10% (vol/vol) FBS and 1% (vol/vol) Penicillin-Streptomycin at 37 °C. Culture MCF-7 cells in DMEM (high glucose) supplemented with 10% (vol/vol) FBS and 1% (vol/vol) Penicillin-Streptomycin at 37 °C. Culture BT474 cells in DMEM (high glucose) supplemented with 5 g $L^{-1}$ glucose, 10% (vol/vol) FBS and 1% (vol/vol) Penicillin-Streptomycin at 37 °C. Culture CEM.$NK^R$ cells in RPMI-1640 supplemented with 10% (vol/vol) FBS at 37 °C. All cells are cultured in 5% $CO_2$ atmosphere.

**U87MG tumor model**   Inject $5 \times 10^6$ U87MG cells subcutaneously into athymic nude mice (on their shoulders). Wait for 3-4 weeks before imaging, until tumor sizes reach 300~500 $mm^3$.

**! CAUTION** Please obtain appropriate training from the institution regarding animal handling and have animal protocols in place before performing animal studies.

**Equipment Setup**

**Raman Microscope** Confocal Raman spectroscopic imaging is carried out using a Horiba-Jobin-Yvon Raman confocal microscope with a 785 nm laser (80mW) as the excitation light source. A 50× objective was used for imaging with ~1 μm laser spot size. A 1 mm pin-pole was applied to restrict the spatial resolution in z-axis to be ~1 μm. Each Raman spectroscopic map contains at least 100 × 100 spectra with a 0.5 s integration time for each spectrum.

**FACS machine** FACS measurement is performed using a Becton Dickinson FACScan. A 488nm laser is used as the excitation light source. Channel one, two and three collect green, yellow and red fluorescence, respectively. A flow rate of 200~400 cells / second is used in the measurement.

# PROCEDURE

**Functionalization of SWNTs**

**1|** Weigh 1mg of Hipco SWNTs (or isotopically modified SWNTs) and 5 mg of PL-PEG5000-Amine or PL-PEG2000-Amine into a 20 ml glass scintillation vial. Add 5 ml of water. Dissolve DSPE-PEG completely by shaking.

**! CAUTION** Hipco SWNTs are very light and could become airborne. Inhalation of SWNTs should be avoided. Wear goggles, lab coat and face-mask during experiments.

**2|** Sonicate the above vial in a bath sonicator for 60 min at room temperature. Change the water in the water bath every 20 min to avoid over heating.

▲ **CRITICAL STEP** Make sure the vial is at the best position in the bath sonicator to ensure the most efficient sonication.**? TROUBLESHOOTING**

**3|** Centrifuge the above SWNT suspension for 6 h, at 24,000 xg, room temperature (~22 °C). Collect the supernatant solution.

**4|** Record the UV-VIS-NIR absorption spectrum of the obtained SWNT solution. The final SWNT concentration normally ranges from 40 mg L$^{-1}$ to 70 mg L$^{-1}$. Store the SWNT solution at 4 °C.

■ **PAUSE POINT** The PL-PEG functionalized SWNT solution can be stored at 4 °C for 1-2 months before bioconjugation. It is highly recommended to store SWNT solutions in the presence of excess PL-PEG (before removal of excess PL-PEG in **Step 5**).

**5|** Add 1 ml of SWNT solution from the stock prepared in Step 3 into a 4 ml Amicon centrifugal filter device with a molecular weight cut off (MWCO) of 100 kDa. Add 3 ml of water and centrifuge the device for 10 min, at 4,000 xg, room temperature. The leftover volume in the filter should be less than 0.5 ml. Fill the filter device with water to 4 ml. Wash 5-6 times by repeating the centrifuge / water adding steps, to completely remove excess PL-PEG in the SWNT solution. Adjust the final SWNT solution to a concentration of ~50 mg L$^{-1}$ by adding the required amount of water. The concentration of SWNT solution is measured by UV-VIS-NIR spectrometer with a weight extinction coefficient of 0.0465 mg L$^{-1}$ cm$^{-1}$ at 808 nm.

**? TROUBLESHOOTING**

**6|** SWNT can be functionalized with Option A Targetting antibodies, Option B radiolabeling, Option C siRNA, Option D with doxorubicin.

**Option A. Conjugation with RGD peptide**

**i)** Dissolve 0.5 mg of Sulfo-SMCC in 50 μl of DMSO. Add 0.5 ml of 50 mg L$^{-1}$ SWNT functionalized by PL-PEG5000-Amine solution from **Step 5**. Add 60 μl of PBS (10×). React at room temperature for 2 h.

**ii)** Remove excess Sulfo-SMCC by an Amicon centrifugal filter device (MWCO = 100kDa). Wash 5-6 times. Add water to a volume of 0.5 ml.

▲ **CRITICAL STEP** The final SWNT solution should be immediately mixed with thiolated RGD.

**iii)** Dissolve ~2 mg of TCEP in 20~25 μl of 0.5 M sodium bicarbonate. Add sodium bicarbonate buffer slowly until the solution reaches pH 6. The pH is measured by a pH test paper. Add required volume (calculated by the exact weight of TCEP and the exact volume of sodium bicarbonate buffer added) of water so that the final TCEP concentration is 0.2 M.

**! CAUTION** Always prepare the TCEP solution immediately before use. TCEP-HCl solution in water has a very acidic pH.

**iv)** Mix 0.5 ml Sulfo-SMCC activated SWNT solution from **Step ii** with 50 μl of 10×PBS, 10 μl of 100 μM RGD-SH solution and 25 μl of the TCEP solution from **Step iii**. The final reaction solution should have a SWNT concentration of ~40 mg L$^{-1}$, RGD-SH concentration of ~0.2 mM and TCEP concentration of ~10 mM. These concentrations are based on the volumes and initial concentrations of various solutions added into the reaction mixture. Allow the reaction to proceed for 24 h at 4 $^o$C.

**v)** Remove excess RGD and TCEP in the above solution using an Amicon centrifugal filter device (MWCO = 100 kDa). Wash 5-6 times as described in Step 5. Store the SWNT-RGD conjugate at ~50 mg L50 mg L$^{-1}$ in water at 4 $^o$C.

■ **PAUSE POINT** The SWNT-RGD conjugate can be stored at 4 $^o$C for 2-3 weeks without losing targeting activity.

**vi)** Cell staining

a) Collect U87MG cells and MCF-7 cells by trypsinization in cell plates and gentle centrifugation for 7 min in 1.5 ml centrifuge tubes, at 300g, room temperature. Re-suspend the pellet in RMPI-1640 cell medium with 10% (vol/vol) FBS. Count the numbers of cells to make sure a density of ~1 million cells ml$^{-1}$..

b) Add 50 μl of SWNT-RGD from **Step v** into 200 μl of U87MG cells (positive) and 200 μl of MCF-7 cells (negative). Incubate for 1 h at 4 °C..

c) Wash the above cells with 200 μl of 1x PBS 3 times after incubation. The cells may be stored in 1x PBS at 4 °C for a few hours before Raman imaging without changing Raman imaging results.

**vii)** Raman Imaging

a) Seal a drop of cell suspension between two thin plastic cover-slips.

b) Carry out Raman spectroscopic imaging of the labeled SWNTs using a Horiba-Jobin-Yvon Raman confocal microscope.[15]

**Option B. Conjugation with Targeting Antibodies**

**i)** Dissolve 0.5 mg of Sulfo-SMCC in 50 μl of DMSO. Add 0.5 ml of SWNT functionalized by PL-PEG5000-Amine solution from **Step 5**. Add 60 μl of PBS (10×). React at room temperature for 2 h.

**ii)** Remove excess Sulfo-SMCC using an Amicon centrifugal filter device (MWCO = 100kDa). Wash 5-6 times as described in Step 5. Add water to a final volume of 0.5 ml.

▲ **CRITICAL STEP** The final SWNT solution should be immediately mixed with thiolated Herceptin.

**iii)** Immediately after the Sulfo-SMCC reaction with SWNTs is started (Option B (ii)), weigh 1~2 mg of Traut's reagent into a 2 ml plastic tube. Based on the weight, add the desired volume of water so that the final concentration of Traut's reagent is 5 mM. Mix 6 μl of 140 μM Herceptin into a 0.5 ml plastic tube with 10 μl of 1×PBS, 1.7 μl of Traut's reagent solution, and 1 μl of 0.5 M EDTA solution. The molar ratio of Traut's reagent:antibody is about 10:1. Incubate the reaction solution for 1.5-2 h at 4 °C.

**iv)** Add 300 μl of 1× PBS into the solution prepared in **Step iii**, remove the excess Traut's reagent by filtration with a Microcon Ultracel YM-100 filter device. Centrifuge for 6~8 min at 10,000 xg, room temperature until the leftover volume is less than 10 μl.

CRITICAL STEP Use the thiolated Herceptin immediately.

**v)** Mix the thiolated Herceptin from **Step iv** with 0.5 ml of Sulfo-SMCC modified SWNTs from **Step ii.** Add 50 μl of 10× PBS and 2 μl of 0.5 M EDTA. Incubate the reaction solution for 24 h at 4 °C. The conjugate can be used directly without further purification.

■ **PAUSE POINT** The SWNT-Herceptin conjugate can be stored at 4 °C for 2-3 weeks without losing targeting activity.

**Option C. Radiolabeling of SWNTs**

**i)** Dissolve 2 mg of DOTA in 50 μl of 0.1 M sodium hydroxide solution. Dissolve 2 mg of Sulfo-NHS and 1.5 mg of EDC in 50 μl of water. Add 30 μl of the Sulfo-NHS/EDC solution to the 50 μl DOTA solution. Incubate for 15 min at room temperature. Check the solution pH using pH test paper. The pH in the incubation solution should be pH 5-6. Slowly add more 0.1M sodium hydroxide solution if the pH is below 5. Molar ratio: DOTA : EDC: Sulfo-NHS ~ 1 : 1 : 1.2.

▲ **CRITICAL STEP** Use EDC and Sulfo-NHS immediately after they are dissolved in water.

**ii)** Dissolve 0.5 mg of Sulfo-SMCC in 20 μl of DMSO and mix with 10 μl of the DOTA:EDC:Sulfo-NHS solution from **Step i**. Add 500 μl of PL-PEG5000-Amine functionalized SWNT solution from **Step 5**. Add 60 μl of 10x PBS. Incubate the reaction solution for 2 h at room temperature.

**iii)** Remove excess Sulfo-SMCC, DOTA, EDC and Sulfo-NHS using an Amicon centrifugal filter device (MWCO = 100 kDa). Wash 5-6 times as described in Step 5. Add water to a final volume of 0.5 ml.

**iv)** Mix 0.5 ml of the SWNT solution from **Step iii** with 50 μl of 10×PBS, 10 μl of 100 μM RGD-SH solution, and 10 μl of 0.2 mM TCEP solution prepared in **Option A, step iii**. The final reaction solution has a SWNT concentration of ~40 mg L$^{-1}$, RGD-SH concentration of ~0.2 mM and TCEP concentration of ~10 mM. Let the reaction proceed for 24 h at 4 $^{o}$C.

**v)** Remove excess RGD and TCEP in the above solution using an Amicon centrifugal filter device (MWCO = 100 kDa). Wash 5-6 times as described in Step 5. Measure the concentration of final SWNT-RGD solution by UV-VIS-NIR spectroscopy. Store the SWNT-RGD conjugate at ~50 mg L$^{-1}$ in water at 4 $^{o}$C.

■ **PAUSE POINT** The SWNT-RGD conjugate can be stored at 4 $^{o}$C for 2-3 weeks without losing targeting activity.

**vi)** Dilute 10~20 mCi of $^{64}$CuCl$_2$ in 0.5 ml of 0.1 M sodium acetate buffer (pH 6.5). Add a fraction of this solution (2 mCi activity) to 100 μl of RGD/DOTA co-conjugated SWNT solution prepared in **Step v**. Add 300 μl of 0.1 M sodium acetate buffer (pH 6.5). Incubate the reaction mixture for 1 h at 40 °C with constant shaking.

**! CAUTION** Please obtain appropriate training for handling radioactive materials. Wear goggles, lab coat, mask, radiation dosimeter badge and rings during experiment. Handling of radioactive isotopes should only be performed in designated rooms for those experiments. Check any possible radioactive contamination after experiments.

**vii)** Remove excess $^{64}$Cu by filtration with a Microcon Ultracel YM-100 filter device. Centrifuge for 6~8 min at 10,000 g until the leftover volume is less than 10 μl. Wash 3~4 times by adding 200-300 μl water and centrifuge for 6~8 min at 10,000 g each time. Re-suspend the final labeled SWNTs in 500 μl of 1X PBS (SWNT concentration ~10 mg L$^{-1}$). Radio labeling yield is determined by γ-counting.

**! CAUTION** Radio isotope wastes including contaminated devices such as filters should be collected and disposed in designated containers shielded with lead.

**? TROUBLESHOOTING**

**viii)** Inject 150 μl of ~10 mg L$^{-1}$ radio labeled SWNTs (**Step vii**) into U87MG tumor bearing mice via the tail vein.

**! CAUTION** Please obtain appropriate training from the institution regarding animal handling. Animal protocols must be in place before performing animal studies.

**viii)** Image the mice under a micro-PET scanner at different time points post injection e.g., 0.5 h, 2 h, 4 h, 6 h, 24 h. Process PET data according to a previously established protocol.[25]

**ix)** Scarifice mice 24 h post injection.

**! CAUTION** The animal bodies should be collected in a designated freezer for radioactive contaminated biohazardous waste. Label all animal cages clearly with radioactive marks. Leave contaminated cages in the radioactive work designated animal room for a week until a complete decay of $^{64}$Cu (half-life ~ 12.7h) before cleaning. Check any possible radioactive contamination after experiments.

**Option D. SiRNA Conjugation to SWNTs via Cleavable Disulfide Bond**

**i)** Mix 500 μl of PL-PEG2000-amine functionalized SWNTs from **Step 5** with 0.5 mg of Sulfo-LC-SPDP. Add 50 μl of 10X PBS. Incubate for 2 h at room temperature.

**ii)** Immediately after **Step i** is initiated, prepare 10 mM of DTT solution by dissolving 1.54 mg of DTT in 1 ml of water. Mix 15 μl of 100 μM siRNA (siRNA$_{CXCR4}$ or siRNA$_{luc}$) with 1.5 μl of DTT solution. Allow the reaction to proceed for 1.5 h at room temperature.

**iii)** After **Step i**, remove the excess Sulfo-LC-SPDP from the SWNT solution using an Amicon centrifugal filter device (MWCO = 100 kDa). Wash 5-6 times by adding 3-4 ml DNAse / RNAse free water and centrifuge for 6~8 min at 10,000 g each time. Keep the final volume to less than 10 μl.

**CRITICAL STEP** The obtained activated SWNTs should be used immediately for siRNA conjugation.

**iv)** After **Step ii**, in parallel with **Step iii**, purify DTT treated with the siRNA using a NAP-5 column following the manufacturer's protocol. Elute the siRNA from the column with DNAse / RNAse free 1X PBS.

**v)** Re-suspend the activated SWNTs from **Step iii** with the 500 μl of purified siRNA solution from **Step iv**. Allow the conjugation to proceed for 24 h at 4 °C. The final SWNT and siRNA concentrations are ~40 mg L$^{-1}$ and ~2.5 μM, respectively.

▲ **CRITICAL STEP** Use the SWNT-siRNA conjugate immediately after synthesis. This will reduce the chance of bacteria contamination and siRNA degradation.

**vi)** For the siRNA transfection, plate CEM.NK$^R$ cells in a 24-well plate with 500 μl of cells per well. The cell density should be ~$1 \times 10^5$ cells ml$^{-1}$.

**vii)** To remove aggregates, centrifuge the SWNT-siRNA solutions prepared in Step v for 10 min, at 10,000 xg, 4 $^o$C. Add 100 μl of the SWNT-siRNA$_{CXCR4}$ or SWNT-siRNA$_{luc}$ conjugate to each well. The final SWNT and siRNA concentrations are ~10 mg L$^{-1}$ and ~500 nM, respectively, in the cell medium.

**viii)** Transfect CEM.NK$^R$ cells with 500 nM siRNA$_{CXCR4}$ using commercial transfection agents, (e.g.,) Lipofectamine2000 (Invitrogen), LipofectamineRNAiMAX (Invitrogen), siPORT (Ambion) and HiPerFect (Qiagen) following the manufacturers' protocols.

**ix)** Incubate the cells at 37 $^o$C, 5% $CO_2$ for 3 d before analysis.
TROUBLESHOOTING

**x)** To analyze the effect of the RNAi, wash each well of cells (non-adherent) with 200 μl 1X PBS twice by centrifuge for 7 minutes, at 300 g, room temperature. Re-suspend the cells in 200 μl of 1X PBS. Add in 2 μl of PE-anti-CXCR4 antibody solution. Incubate at for 1 h at 4 $^o$C.

**xi)** Wash the cells with 200 μl 1X PBS 3 times by centrifuge for 7 minutes, at 300 g, room temperature. Dilute 5 μl of 1 mg ml$^{-1}$ PI solution in 5 ml of 1X PBS. Re-suspend the cells in 200 μl of the PBS containing the 1 μg ml$^{-1}$ PI. Store the cells at 4 $^o$C prior to FACS analysis.
**PAUSE POINT** The stained cells can be stored at 4 $^o$C for up to 4 hours without losing much viability.

**xii)** Cells analysis by flow cytometry. Measure PE and PI fluorescence by channel 2 (yellow) and 3 (red), respectively. Exclude PI positive cells (dead cells) in the data analysis. Determine the CXCR4 expression levels on CEM.NK cells after various treatments by the meaning fluorescence intensity of each cell sample.

**? TROUBLESHOOTING**

**Option E. Doxorubicin Loading on Functionalized SWNTs**

**i)** Mix 0.5 ml of PL-PEG5000-Amine functionalized SWNTs from **Step 5** or RGD conjgauted SWNTs from **Option A, step v** with 30 μl 10 mM DOX solution. Add 50 μl 10xPBS and 2 μl 0.5 M sodium bicarbonate buffer. The final pH is around 8. The final SWNT and DOX concentrations are ~40 mg/L and ~0.5 mM, respectively. Incubate the solution at 4 °C for 24 hours.

**▲ CRITICAL STEP** Make sure the pH is not too basic. The SWNT-RGD-DOX will have reduced stability if pH is over 9.

**ii)** Remove excess DOX by an Amicon centrifugal filter device (MWCO = 10kDa). Wash at least 6 times with 3-4 ml water each time until the filtrate solution appears to be almost colorless. Re-suspend SWNT-DOX in 0.5 ml of water. Centrifuge the SWNT-DOX solution for 10 min, at 10,000 xg, 4 °C, to remove aggregates.

**? TROUBLESHOOTING**

**iii)** Record UV-VIS-NIR absorption spectra of SWNT (SWNT-RGD) and SWNT-DOX (SWNT-RGD-DOX) conjugates.[21] To determine DOX loading onto SWNTs normalize the spectra by the absorption at 808 nm. DOX has a molar extinction coefficient of $1.05 \times 10^4$ $M^{-1} \cdot cm^{-1}$. The final DOX loading should be 600-800 DOX molecules per SWNT.

■ **PAUSE POINT** The SWNT-DOX conjugate can be stored at 4 °C for 3~5 days without obvious losing of activity.

**iv)** To determine the cell toxicity of the DOX conjugated SWNTs plate U87MG and MCF-7 cells in two 96-well plates with ~10,000 cells in 100 μl medium per well. Culture the cells overnight at 37 °C, 5% $CO_2$.

**v)** Centrifuge the SWNT-DOX solutions prepared in Step (iii) for 10 min at 10,000 xg, 4 °C, to remove any aggregates. Add a series of concentrations (1 μM to 40 μM) of free DOX, SWNT-DOX and SWNT-RGD-DOX into different wells of cells. Untreated cells in other wells are used as the control. Add each sample to triplicate wells. Incubate the plates for 1-2 h at 37 °C, 5% $CO_2$.

**vi)** Gently wash cells with 200 μl 1x PBS twice for 7 minutes at 300g, room temperature, to remove excess drug. Add 100 μl of fresh cell medium and incubate at 37 °C, 5% $CO_2$ for 48 h.

**vii)** Measure the cell viability in each well using a CellTiter-96 kit following the manufacturer's instructions. Measure the absorbance of the cells at 490 nm and determine the relative cell viabilities.[21]

## TIMING

Step 1-5, functionalization of SWNTs 1-2 d

Step 6 Option A, Conjugation of SWNTs with RGD peptide 3 d

Step 6 Option B, Conjugation of SWNTs with antibody (Herceptin) 2 d

Step 6 Option C, Radiolabeling of SWNTs 5 d

Step 6 Option D, siRNA conjugation on SWNTs 5-6 d

Step 6 Option E, Doxorubicin loading on functionalized SWNTs 5-6 d

# TROUBLESHOOTING advice is showin in Table 1

## Table 1 TROUBLESHOOTING Advice

| Step | Problem | Possible Reason | Solution |
|---|---|---|---|
| 2 | SWNTs are not well suspended by PL-PEG. | Sonication is not efficient due to inappropriate vial position in the sonicator. | Adjust the position and angle of the vial in the sonicator. Make sure the water bath is at the recommended level. |
| 5 | Nanotubes are very sticky on the filter device during filtration.. | A certain amount of stickiness is normal for PL-PEG2000- Amine functionalization but not for PL-PEG5000- Amine This maybe due to poor coating of PL-PEG molecules on nanotubes. 1) The PL-PEG/SWNT ratio is not enough when making nanotube suspensions. 2) Water bath is too hot during sonication. | 1) Reduce the amount of SWNTs or increase PL-PEG concentration during sonication. 2) Change water in bath sonicator more frequently during sonication. |
| Option C, vii | The radiolabeling yield is very low (e.g. <30%). | This maybe due to metal ion contamination in the SWNT solution. | Dialyze the SWNT solution against deionized distilled water with a 10kDa MWCO membrane for 2 d before radiolabeling. |
| Option D, ix | Cells are contaminated by bacteria after 3 days incubation. | SWNT solution is not well sterilized before adding into the cell culture. | Normally centrifuging SWNT-siRNA solution for 10 minat 10,000 g will resolve this problem. Repeating the centrifuge step 2-3 times will be helpful. Carefully take out supernatant and discard any aggregates after centrifugation. Use sterilized containers during experiments. |

| | | |
|---|---|---|
| **Option E, ii** | A large amount of SWNT-DOX complex precipitates during centrifugation. | The DOX loading on nanotubes is too high in this case. This stability issue is more frequent when making SWNT-RGD-DOX. | 1) Reduce pH in the loading solution to 7.8-8.0 by adding less sodium bicarbonate buffer. 2) Add 10~20 μM PL-PEG into the loading solution. This will increase the stability of SWNT-RGD-DOX and not affect DOX loading and RGD targeting. |

## ANTICIPATED RESULTS

**Functionalization of SWNTs**

After **Step 5,** PL-PEG functionalized SWNTs should have excellent water solubility and are stable in various biological solutions without any visible aggregation after a long period of incubation time (weeks) (Fig 5). Atomic force microscope (AFM) images of SWNTs show that the lengths of SWNTs range from 50 nm to 300 nm, with an average of ~100 nm.

**Targeting SWNT bioconjugates for Raman imaging and sensing**

In **Option A, step vii,** confocal Raman spectroscopic images of cells stained by SWNT-RGD should exhibit strong SWNT Raman signals on integrin $α_vβ_3$ positive U87MG cells but not on negative MCF-7 cells (Fig. 6a). Non-specific binding of SWNTs to MCF-7 cells should be minimal. The ratio of Raman signals on positive *verus* negative cells is higher than 40 (Fig. 6b). Raman imaging of BT474 (Her2 positive) and MCF-7 (Her2 negative) cells after staining with SWNT-Herceptin (**Option B**) should give very similar results (data not shown). Those targeted SWNT bioconjugates may also be used in ultra-sensitive protein microarray detection.[14]

**Radiolabeled SWNTs for in vivo PET imaging and tumor targeting**

In **Option C, step viii,** PET images (Fig. 7) at 6 hours post injection should show high tumor uptake (10-15%ID/g) in U87MG tumor bearing mice injected with SWNT-RGD (functionalized by PL-PEG5000).[25] Radiolabeled SWNTs without RGD conjugation should show a reduced uptake in the tumor (3-5%ID/g) in comparison with RGD conjugated SWNTs.. Tumor uptake will be significantly reduced if mice are pre-injected with a high dose of free RGD peptide before injection of SWNT-RGD (4-6%ID/g). Control integrin $\alpha_v\beta_3$ negative HT29 tumors should have a lower uptake of SWNT-RGD (3-5%ID/g) (Fig. 7d).

**SWNT based siRNA transfection and RNAi effect**

In **Option D, step vii,** as shown in Fig. 8, SWNT only and SWNT-siRNA$_{luc}$ mis-matched control treated CEM.NK$^R$ cells show normal CXCR4 expressions.[4] CXCR4 expression should be significantly reduced after treatment with SWNT-siRNA$_{CXCR4}$ (two sequences) compared to untreated cells. The RNAi effect should range from 70-90%. Other types of commercial cationic liposome based siRNA transfection agents do not show significant siRNA transfection effects to CEM.NK$^R$ cells because human T cells are well known hard-to-transfect cells.

**Doxorubicin loading on functionalized SWNTs**

In **Option E, the** SWNT-DOX solution should show a reddish color due to the UV-VIS absorption of DOX stacked onto SWNTs (Fig. 9a). The DOX absorption peak at 490 nm after subtraction of SWNT absorption at this wavelength is used to determine DOX concentration (Fig. 9b). In **Option E, step iv**, SWNT-DOX has a lower toxicity than free DOX. Conjugation of RGD enhances the toxicity of DOX loaded SWNTs to integrin $\alpha_v\beta_3$ positive U87MG cells but not to negative MCF-7 cells (Fig. 9 c&d).[21]

## ACKNOWLEDGMENTS

The multiple projects involved here were supported by a Stanford Graduate Fellowship, a Stanford Bio-X grant, CCNE-TR at Stanford University, NIH-NCI R01 CA135109-02 and Ensysce Biosciences Inc. Drs. Nadine Wong Shi Kam, Sarunya Bangsaruntip, Xiaowu Tang, Xiaoming Sun, Xiaoyuan Chen, Weibo Cai, and Ms. Nozomi Nakayama have also contributed in the development of this protocol.

**CFI**

The authors declare that they have no competing financial interests.

**AUTHOR CONTRIBUTIONS STATEMENT**

Z.L and H.D designed and wrote this paper. S.T and C.Z revised the paper.

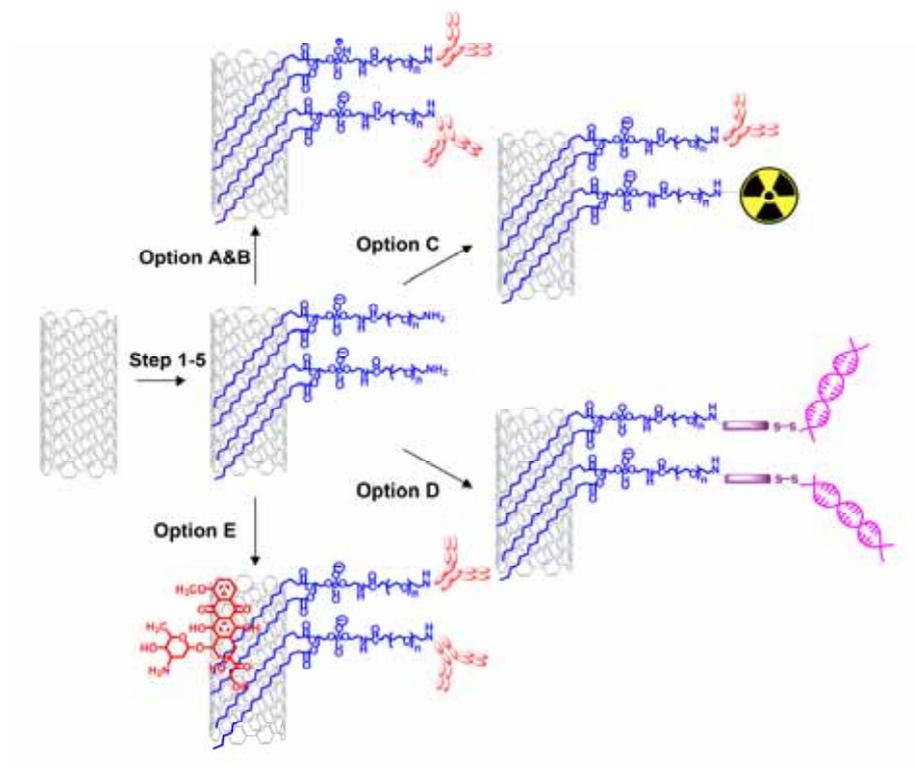

**Figure 1.** Overview of the protocol. The protocol presented here contains five sub-protocols. Step 1-5 is the functionalization of SWNTs. Options A and B are conjugation of targeting ligands to SWNTs. Option C is radiolabeling of SWNTs. Option D is siRNA conjugation to SWNTs. Option E is doxorubicin loading onto functionalized SWNTs.

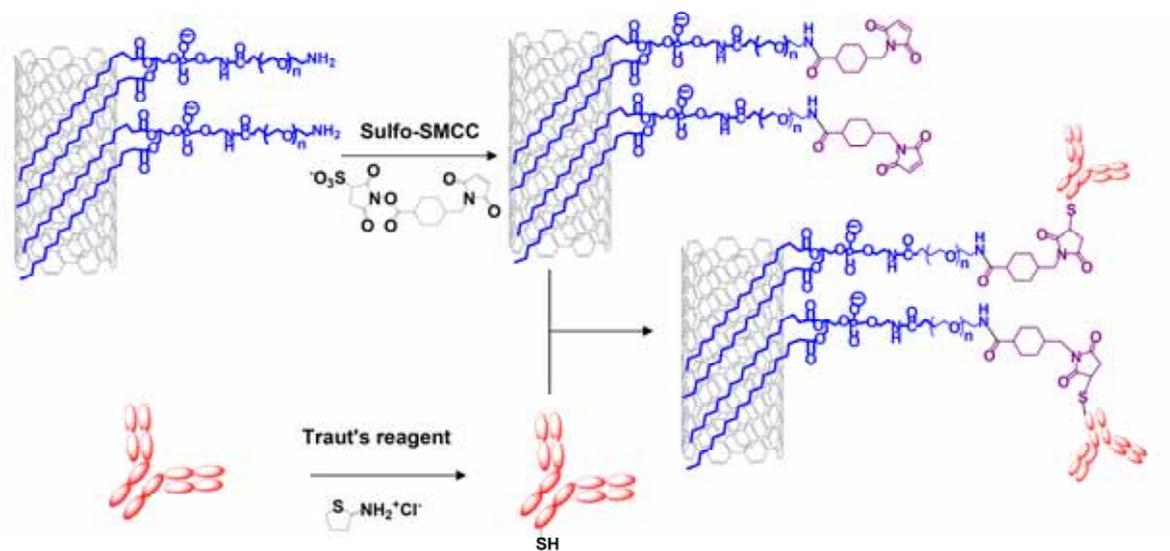

**Figure 2.** A scheme showing conjugation of targeting ligands to SWNTs. PL-PEG5000-Amine functionalized SWNTs are first activated by Sulfo-SMCC, yielding maleimide groups on SWNTs available for conjugation to thiolated antibodies or peptides. Thiolation of antibodies is carried out by treating them with Traut's reagent. Thiolated RGD peptide is synthesized following a previous protocol and used directly.[32]

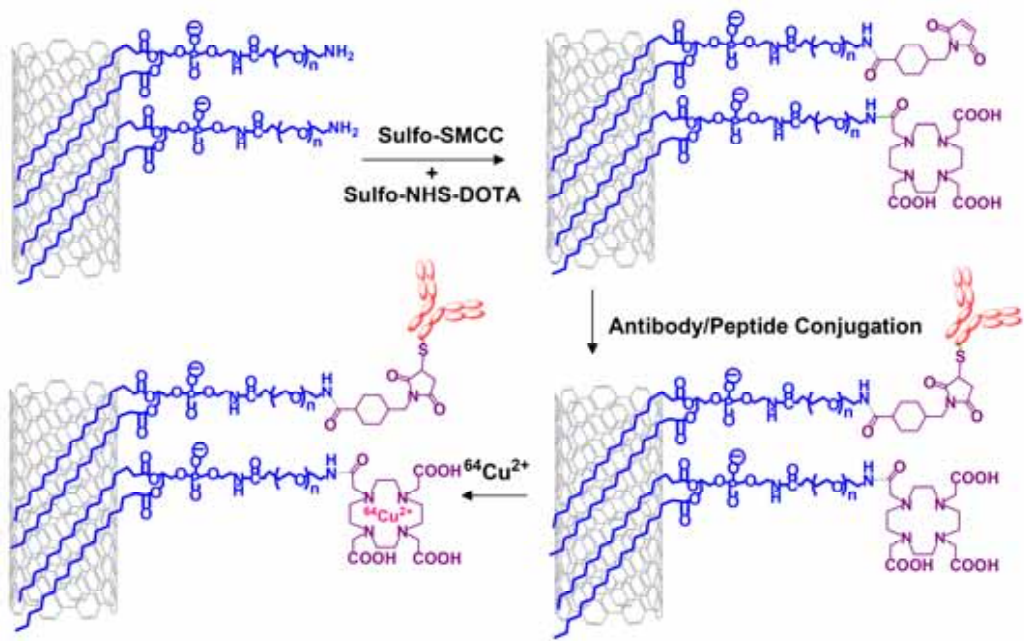

**Figure 3.** A scheme showing radiolabeling of targeting SWNTs bioconjugates. PL-PEG5000-Amine functionalized SWNTs are reacted with a mixture of Sulfo-SMCC and Sulfo-NHS activated DOTA. Thiolated peptides or antibodies are then conjugated to maleimide groups on SWNTs. The targeting SWNT bioconjugate is then labeled by radio isotopes such as $^{64}$Cu through DOTA capture of metal ions.

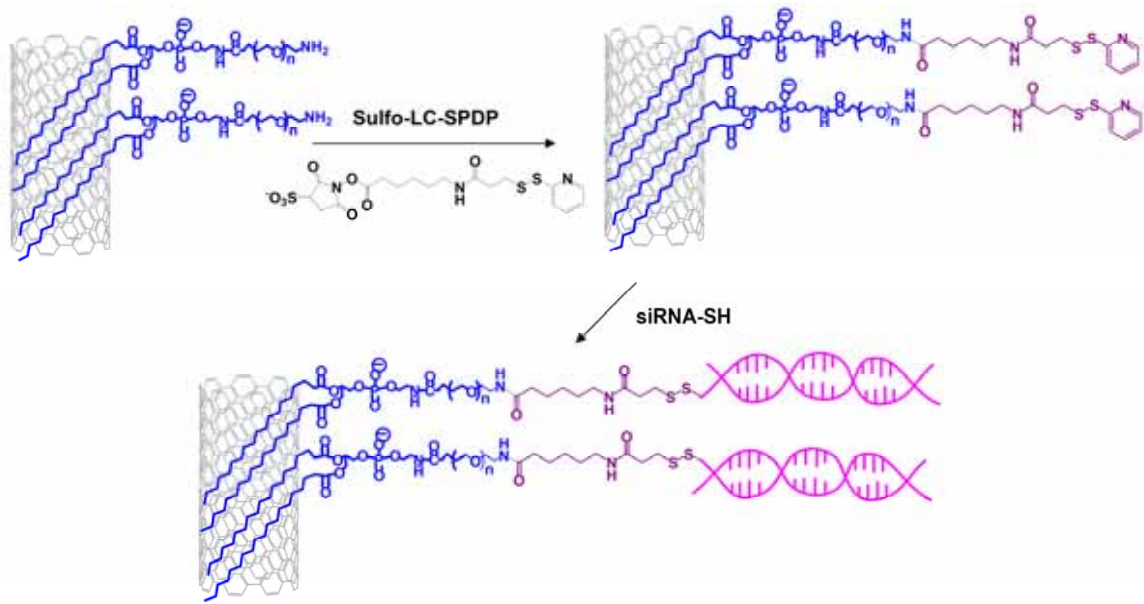

**Figure 4.** A scheme showing siRNA conjugation to SWNTs via a disulfide bond. PL-PEG2000-Amine functionalized SWNTs are activated by the Sulfo-LC-SPDP bifuncitonal linker. The pyridyl disulfide group can then be coupled to thiolated siRNA to create a disulfide linkage via a thiol exchange reaction.

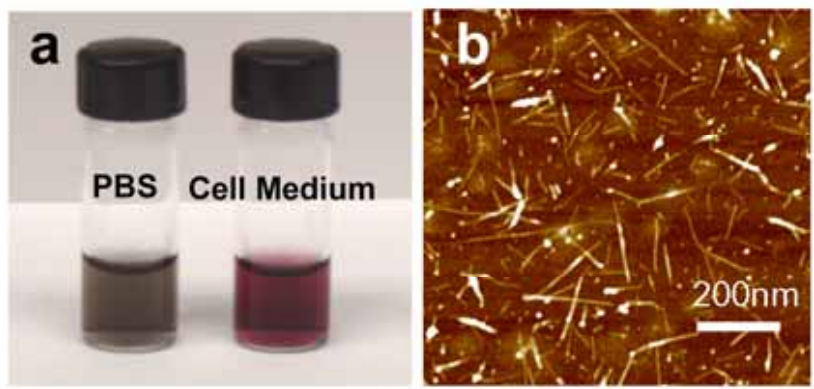

**Figure 5.** Functionalization of SWNTs by PL-PEG. **(a)** A photo showing PL-PEG2000-Amine functionalized SWNTs in PBS and RPMI-1640 cell medium (10% vol/vol FBS). The non-covalently functionalized SWNTs exhibit excellent stability in various biological solutions even after removal of excess PL-PEG molecules from the solution. **(b)** An Atomic Force Microscopy image of SWNTs on a silicon substrate. These SWNTs are mostly single nanotubes with a few small bundles, showing lengths of 50-300 nm. This figure is adapted from our previously published work with copyright obtained from Wiley InterScience.[4]

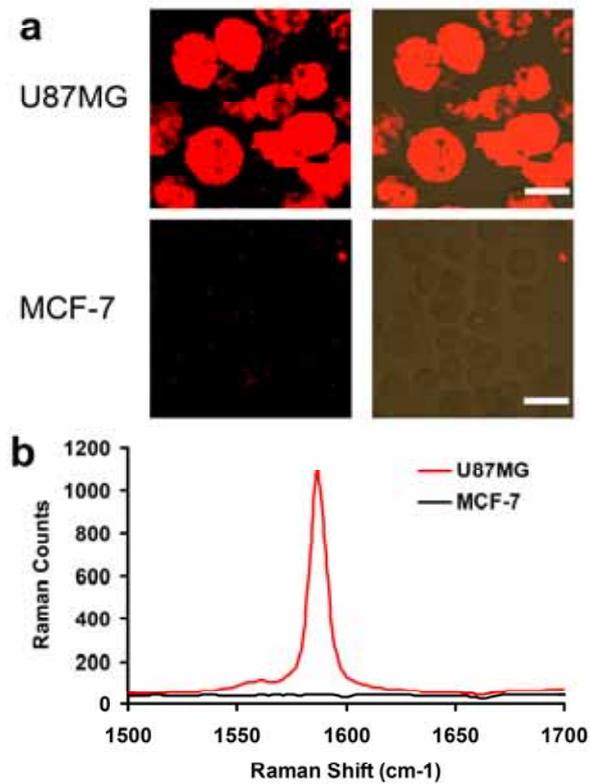

**Figure 6.** Targeting SWNT bioconjugates for cell labeling and Raman imaging. Confocal Raman images of U87MG (integrin $\alpha_v\beta_3$ positive) and MCF-7 (integrin $\alpha_v\beta_3$ negative) cells (Left column) after incubation with SWNT-RGD for 1 hour at 4 °C **(a)**. Optical images overlaid with Raman images are shown in the right column. Scale bar = 20 μm. **(b)** Averaged spectra of two Raman images in **(a)**. Very strong Raman G-band signals are observed on U87MG cells. Negative MCF-7 cells exhibit minimal non-specific absorption signals of SWNT-RGD. This figure is adapted from previously published work[14]

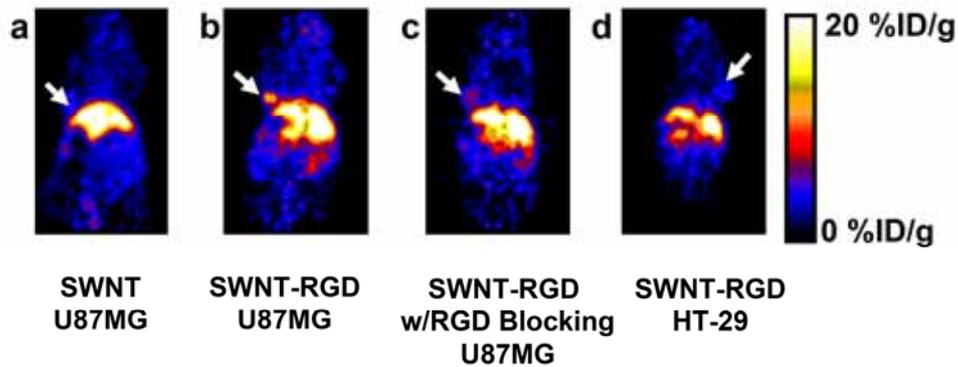

**Figure 7.** Radiolabeled SWNTs for *in vivo* PET imaging and tumor targeting. Images are obtained at 6 hours post injection of radiolabeld nanotubes. **(a)** Control SWNTs without RGD conjugation show low uptake in the tumor. **(b)** High tumor SWNT uptake is observed for mice injected with SWNT-RGD. **(c)** A control experiment showing blocking of SWNT-RGD tumor uptake by co-injection of free c(RGDyK). **(d)** A control experiment showing low uptake of SWNT-RGD in integrin $\alpha_v\beta_3$-negative a tumor formed from HT-29 cells. Efficient tumor targeting is achieved by conjugating SWNTs with RGD peptide, which binds specifically to integrin $\alpha_v\beta_3$ expressed on tumor cells and tumor vasculature. The arrow points to the tumor on the mouse. Animal experiments were conducted under the protocols of *Administrative Panel on Laboratory Animal Care* (APLAC) at Stanford University. This figure is adapted from our previously published work[25]

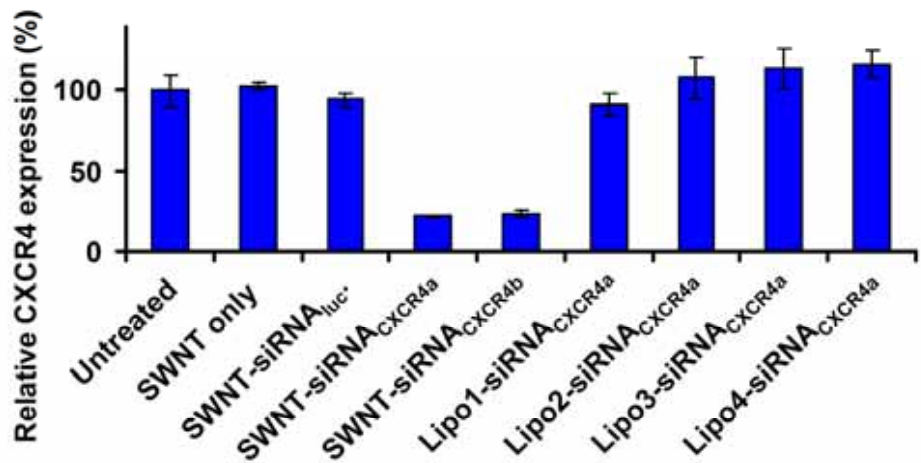

**Figure 8.** CXCR4 expression levels on CEM.NK cells after various treatments indicated, including SWNT-siRNA$_{CXCR4a}$, SWNT-siRNA$_{CXCR4b}$, four types of commercial cationic liposome agents (Lipo1-4) and luciferase (Luc) siRNA control (SWNT-siRNA$_{luc}$). The siRNA concentration is 500 nM in the incubation solution. Cells are incubated for 3 days prior to the FACS analysis. The expression levels of CXCR4 are determined using FACS analysis. Human T cells are not transfected by liposome based transfection agents. In contrast, SWNT based siRNA transfection method appears to be highly efficient to those cells. The error bars are based on standard deviation (SD) of triplicated samples. siRNA$_{CXCR4a}$ and siRNA$_{CXCR4b}$ both target CXCR4 mRNA but at different locations (different sequence). This figure is adapted from our previously published work with copyright obtained from Wiley InterScience. [4]

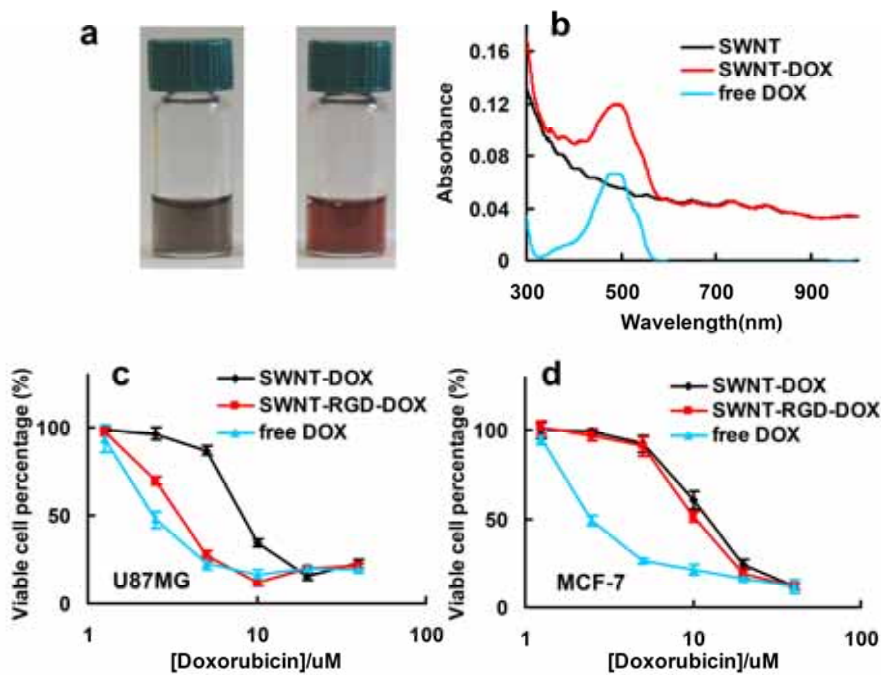

**Figure 9**. Doxorubicin on functionalized SWNTs for drug delivery. **(a)** Photos of PL-PEG functionalized SWNT solutions with and without bound doxorubicin. **(b)** UV-VIS-NIR absorbance spectra of solutions of free doxorubicin (light blue), SWNTs with PL-PEG functionalization (black), and PL-PEG SWNTs complexed with doxorubicin (red) after simple incubation in a doxorubicin solution at pH 8-9. The absorption peak at 490nm is due to doxorubicin π-stacked on SWNTs, and used for analyzing the amount of molecule loaded onto nanotubes. **(c&d)** Concentration dependent survival curves of U87MG cells **(c)** and MCF-7 cells **(d)** treated with the various samples indicated. The viable cell percentages are measured by the MTS assay (CellTiter 96 kit). SWNT-DOX has relatively lower toxic effect than free DOX to both types of cells while SWNT-RGD-DOX exhibits increased toxicity to U87MG cells but not to MCF-7 cells. The viable percentage is normalized to the untreated control sample, which is determined as 100%. Error bars are based on standard deviations of triplicated samples. This figure is adapted from our previously published work with copyright obtained from ACS Publications.[21]